\documentclass[journal, a4paper]{IEEEtran}
\IEEEoverridecommandlockouts
\usepackage{booktabs}
\usepackage{cite}
\usepackage{amsmath,amssymb,amsfonts}
\usepackage{algorithmic}
\usepackage{graphicx}
\usepackage{textcomp}
\usepackage{url}
\usepackage{xcolor}
\usepackage{tablefootnote}

\newcommand{\FigRef}[1]{Fig.~\ref{#1}}

\newcommand{\SecRef}[1]{Section~\ref{#1}}
\newcommand{\FA}{\mathit{FA}}
\newcommand{\fa}{\mathit{fa}}
\newcommand{\textapprox}{\raisebox{0.5ex}{\texttildelow}}

\begin{document}

\title{Implementation Considerations for ACAS and Simulation Results}

\author{Jón Winkel,
      Ignacio Fernandez-Hernandez,
      Cillian O'Driscoll~\IEEEmembership{Member,~IEEE} }

\maketitle
\begin{abstract}
      The Assisted Commercial Authentication Service (ACAS) is a semi-assisted signal authentication concept currently being defined for Galileo, based on the E6-C encrypted signal. Leveraging the assumption that the true E6-C encrypted signal always arrives before any inauthentic signal, we define user concepts for signal detection, including vestigial signal search. We define three mitigation levels, each level defending against an increasing set of threats, incorporating the described concepts and additional checks. The concepts are analyzed and implemented in a simulation environment, and tested in both nominal conditions and under advanced spoofing attacks. The results suggest that even advanced attacks can be detected and mitigated by ACAS receivers.
\end{abstract}


\section{Introduction\label{introduction}}

Global Navigation Satellite Systems (GNSS) location is based on two main inputs: positions and time information of the satellites, and measurements (pseudorange, carrier phase etc.) obtained at the user receiver. Galileo OSNMA already provides authentication of the data \cite{fernandez2016navigation,EUSPA_OSNMA}, possibly with other systems following suit in the future, such as GPS's CHIMERA \cite{Anderson2017}.

The measurements are more difficult to authenticate, as they are generated in the receiver. The two main concepts currently under consideration are encryption of the spreading code using symmetric encryption, and encryption of the spreading code (or parts thereof) using delayed key (or signal) disclosure, as is the case for CHIMERA \cite{Anderson2017} or OSNMA, at data level \cite{fernandez2021analysis}. Both schemes allow the user to (mostly) apply the crucial assumption that the encrypted authentic signal will always arrive \emph{before} any inauthentic signal.

In the case of symmetric encryption, the keys must be kept secret, which is possible for governmental services (e.g., Galileo PRS). However, for open services this is usually not considered feasible and a delayed disclosure scheme is preferred. A key property of the delayed disclosure scheme is that the user must know for certain that the signal was received \emph{before} the key was disclosed. This places requirements on absolute time synchronization on the user, and inherently introduces a latency between the signal reception and the availability of the PVT.

The delayed disclosure can be implemented in several forms. Ideally, the stream of keys that were used to encrypt the signal is distributed to the users with appropriate latency \cite{scott2003anti}. Alternatively, the recorded samples can be transmitted to a trusted server with access to the keys or the signal, which then returns the PVT solution back to the user\cite{Lo2009}. The encrypted code chips themselves can also be transmitted to the user via a trusted side channel after signal in space transmission. In the case of ACAS (Assisted Commercial Authentication Service), yet another approach is introduced: The symmetrically encrypted signal on Galileo E6-C is re-encrypted using the OSNMA keys (so-called re-encrypted code sequences or RECS) and made publicly available for download up to a week before the encrypted E6 signal is transmitted via SIS. Downloading these in advance, the user can decrypt the RECS, once the OSNMA keys are received, obtaining the encrypted E6 code chips, ready for correlation. The ACAS concept has already been prototyped\cite{fernandez2022semi,terris2022operating,TerrisGallego2023}, it is being studied for various use cases \cite{ardizzon2022authenticated}, and it is expected to start an initial capability once the Galileo E6-C signal is encrypted.

In this paper we first present a number of assumptions for the ACAS processing and the receiver architecture, and then define a few user level algorithms that we consider essential for good protection using ACAS. To accommodate different levels of security requirements users may have, we define three mitigation levels. Finally we present simulation results, demonstrating the performance of an ACAS enabled receiver to both detect inauthentic signals as well as even maintaining operation during an advanced spoofing attack.

\section{ACAS Assumptions and Errors Sources\label{sec:ACAS_assumptions}}

Before starting the analysis we list and discuss some important general assumptions as well as some assumptions on the receiver architecture.

\textbf{Assumption 0:} It is assumed that the basic requirements for ACAS
are fulfilled\cite{FernandezHernandez2023}

\textbf{Assumption 1:} It is assumed throughout that the cryptography remains intact. That is, only attacks that are unable to break the cryptographic material are considered.

\textbf{Assumption 2:} The true encrypted signal in E6 is assumed to always be available, and to always reach the victim \textbf{\emph{before}} its inauthentic replica.

Assumption 2 only strictly holds if the true signal is received directly. If the receiver tracks the signal via a multipath, then it is possible that a re-radiated signal reaches the victim before the true signal. We do not consider the reflected-only case of the true signal to be a viable attack, since only in very rare and constructed cases can the attacker rely on this signal configuration.

The assumption on the availability of the encrypted signal implies that an attacker is unable to null the encrypted signal and that it is not being blocked.

\subsection{Assumptions on the Receiver Architecture\label{assumptions-on-the-receiver-architecture}}

The ACAS receiver implementation is expected to follow a number of assumptions for the design:

\begin{enumerate}
      \item Full tracking on E1 is available continuously and pseudoranges, carrier phases and Doppler values are available.
      \item The carrier phase measurements are used to estimate the range rate over the integration interval of the samples.
      \item The front-end design is assumed to be compatible with a state-of-the-art dual frequency implementation on E1 and E6. In particular:
            \begin{enumerate}
                  \item Both front-ends are assumed to be driven by the same oscillator.
                  \item The front-ends are synchronized such that the sample count of a sample acquired by one front-end can be associated with the sample count of a corresponding sample acquired on the other front-end.
                  \item All digital signal conditioning is assumed to be stable and constant and as such calibratable. This is likely to be the case for static filtering. However, if filters are dynamically turned on and off, for example to filter interference, then this must be accounted for.
                  \item The only non-calibratable delay difference is assumed to come from analog components and sub-sample thermal variations of the ADC. Depending on the ADC, this can easily become in the order of a nanosecond.
            \end{enumerate}
\end{enumerate}

\subsection{Basic ACAS Processing and Receiver Errors\label{basic-acas-processing-and-receiver-errors}}

We propose to start from the E1 measurements and to hand over from measurements performed in E1 over to E6. Since most of the errors on the pseudorange are common on both E1 and E6, and because the timing of the RECS is essentially the time of transmission, the hand-over can be achieved in a relatively straight-forward manner from the definition of the pseudorange: $P_r := ct_{rx} -ct^{tx} \Rightarrow t^{tx} = t_{rx}-P_r/c$. The only remaining error sources are:

\begin{itemize}
      \item Thermal noise: Present both in the E6 samples and on the E1 tracking process, and typically in the order of 0.5-1 m. \cite{J.W.Betz2000c}
      \item Broadcast Group Delay (BGD) between E1 and E6: This is accounted for by the BGD files of the ACAS protocol \cite{FernandezHernandez2023}.
      \item Dispersive atmospheric effects (differential ionosphere error): The residual, differential ionospheric error can be in the order of 30 ns (10 m)\cite{RoviraGarcia2015,RoviraGarcia2019}.
      \item Non-calibrated delay differences in the receiver (analogous to the BGD in the satellite): HW offset between E6 and E1: The main contributor is typically the digital processing. However, this is constant and can be completely compensated. The residual analog component is typically smaller than 10 ns (3 m).
      \item Multipath: The differential multipath error between E1 and E6 can be in the order of 30 ns (10 m).
\end{itemize}

The magnitude of the errors mentioned above are just typical orders of magnitude. The error modelling will be treated in more detail in sec.~\ref{sec:ACAS_Modelling}.

Provided the assumptions previously stated hold, everything else (receiver clock, satellite clock, ephemeris, troposphere etc.) cancels out, since it affects the measurements on both frequencies in the same way \cite{TerrisGallego2023}.

When logging the samples from E6 and the measurements on E1, the sample batch duration must be chosen such that it contains signals from all satellites, accommodating the un-predictable RECS offsets\cite{FernandezHernandez2023}. The receiver should log the measurements (pseudorange, phase, Doppler, Doppler rate) with a sufficient data rate, such that the timing and frequency offsets of the E6 signals can be interpolated with sufficient accuracy. This depends on the expected dynamics on the line-of-sight and clock stability.

If the assumptions in the receiver architecture above are fulfilled, a very limited signal search (\textapprox1-10 code offsets and no Doppler search) is required in E6.

\section{ACAS User Level Processing\label{sec:ACAS_user_algorithms}}

This section describes user algorithms that could be implemented in a receiver. The emphasis is on the signal processing and it is assumed that the cryptographic operations described in \cite{FernandezHernandez2023} are implemented.

In general, the basic ACAS processing can be separated into three steps:

\begin{enumerate}
      \item Detect the presence of the encrypted E6 signal.
      \item Estimate an authenticated E6 range either directly from the E6 snapshot or extrapolating to the E1 measurements.
      \item A position based on measurements from the previous step and authenticated data from OSNMA.
\end{enumerate}

To improve protection additional measures are necessary, which will be discussed further in sec.~\ref{sec:ACAS_mitigation}.

In this paper we define the following ACAS user algorithms:

\begin{itemize}
      \item Signal Detection: When detecting the signal in E6, at least two strategies can be pursued: selecting the correlator with the maximum power, or select the earliest correlator with enough power to exceed the detection threshold. Both possibilities will be analyzed here.
      \item Estimation of the E6 code phase through linear interpolation of the early, late and punctual correlators.
      \item Vestigial Signal Search (VSS): If a vestigial signal is found in E1, hand over to E6 and verify its presence in E6 and select the earliest signal to be the true signal. This is compared to an exhaustive search in E6.
\end{itemize}

\subsection{Maximum Signal Detection\label{maximum-signal-detection}}

A simple detection algorithm for detecting the ACAS signal is to scan a certain range of correlation offsets and chose the maximum signal, as follows:

\begin{enumerate}
      \item Hand-over of the timing and frequency from E1 to E6, extracting the timing of the signal from the E1 measurements.
      \item Correct for inter-frequency effects, mainly receiver calibration of inter-front-end delays, broadcast group delay (BGD), and ionosphere.
      \item Perform a coherent integration over the entire RECS for a selected number of code offsets (\textapprox1-10 offsets). In most cases, one frequency bin should be sufficient, as the frequency offset can be estimated from the E1 carrier phase or Doppler.
      \item Calculate the detection threshold, depending on desired false alarm probability ($P_{\fa}$) (see e.g., \cite{Kaplan2006}).
      \item Check that the expected carrier-to-noise density ratio $C/N_0$ (inferred from E1 and antenna and front-end calibration data) gives rise to the expected minimum probability of detection ($P_d$).
      \item Select the correlator with the highest power, which is also above the detection threshold.
\end{enumerate}

This algorithm is sometimes assumed in the literature\cite{Psiaki2016}. It will simply find the strongest signal that is also above the detection threshold. If a spoofing signal is present, it is to be expected that this algorithm will happily capture the spoofing signal, if it is just slightly stronger than the true signal.

\subsection{Early Signal Detection\label{early-signal-detection}}

The previous maximum signal detection algorithm does not leverage assumption 2, i.e., that the true signal is always the earliest signal. Therefore, we consider it a better strategy to actively look for and identify the earliest signal.

Note that this algorithm tries to deal with the presence of a close-in spoofing signal. The goal is to identify the true signal and separate it from a potential spoofing signal. This is important for the s-curve based range estimate to function properly, which will be discussed in the next section.

Note also that a close-in spoofing signal that is within a few chips or even a fraction of a chip implies that the spoofing signal is only delayed by a few microseconds. This is a very difficult attack to implement in general and places severe restrictions on the attacker (e.g.~the attacker must be very close to the victim.). However, in specific cases, it may be feasible to mount such an attack. See the discussion in sec.~\ref{sec:ACAS_assumptions}.

The early signal detection algorithm proposed, assumes the availability of the magnitude of multiple correlators (or sequential evaluations of one correlator) evaluated at equidistant offsets, arranged in a vector, $\vec{c}$, with $N$ elements.

Increasing the correlator spacing can be useful when the signal is noisy, so the correlator spacing used for the s-curve can be larger than the distance between adjacent elements i.e., an integer $n \ge 1$. The correlator values involved for the s-curve calculations are then $c_{i-n}$, $c_i$ and $c_{i+n}$. The steps of the algorithm are as follows:

\begin{enumerate}
      \item Perform steps 1 to 5 of the maximum detection algorithm above.
      \item Scan the correlators ($c_n, c_{n+1}, \ldots ,c_{N-n}$) from ``early'' to ``late'' and select the first punctual correlator (i.e. $c_i$) that exceeds the acquisition threshold.
      \item If the late correlator ($c_{i+n}$) has a higher value than the punctual one ($c_{i+n}>c_{i}$), increment the index of all three correlators by \emph{one} ($i \rightarrow i+1$). Repeat until the local maximum has been identified.
      \item Evaluate $D_i:= |c_{i+n} - c_{i-n}|$ and minimize $D_i$ under the constraint that $c_i > c_{i+n} \wedge c_i > c_{i-n}$.
      \item If no set of correlators $c_{i-n}$, $c_i$ and $c_{i+n}$ within the vector $\vec{c}$ can be found that meet the criteria above, the signal is not detected.
\end{enumerate}

\begin{figure}[!t]
      \centering
      \includegraphics[width=8cm]{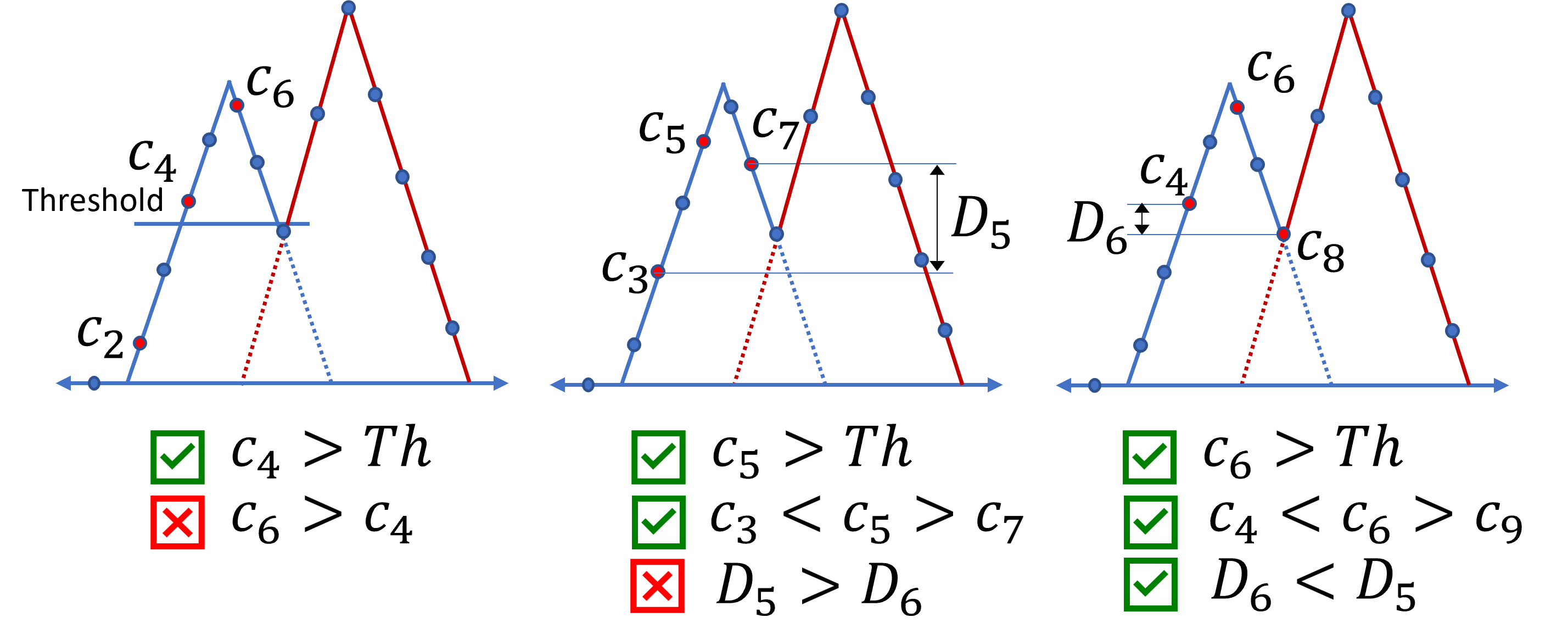}
      \caption{Early signal detection algorithm. The blue triangle on the left corresponds to the early true signal and the red triangle corresponds to the inauthentic signal.
            \label{fig:early_peak_det}}
\end{figure}

The steps in the algorithm are illustrated in the figure \ref{fig:early_peak_det} below for the case where the correlator spacing is 2 ($n=2$). In the first picture, the middle correlator ($c_4$) is above threshold, but it is not the largest one. In the second picture, the middle correlator ($c_5$) is above the threshold and it is larger than the early ($c_3$) and late ($c_7$) correlators. However, it is not the optimal position. Moving one correlator to the right ($c_5 \rightarrow c_6$) will maximize the symmetry of the correlator configuration ($D_6 < D_5$).

\subsection{Range Bias Detection (S-Curve Root-Finding)\label{sec:root_finding}}

When the signal has been detected, using either of the two detection algorithms above, the range offset between E1 and E6 can be calculated. There are several ways to do this. For simplicity we select the linear interpolation defined in\cite{Risueno2005}. Other methods are described in \cite{borre2022gnss}.

\begin{figure}[!t]
      \centering
      \includegraphics[width=6cm]{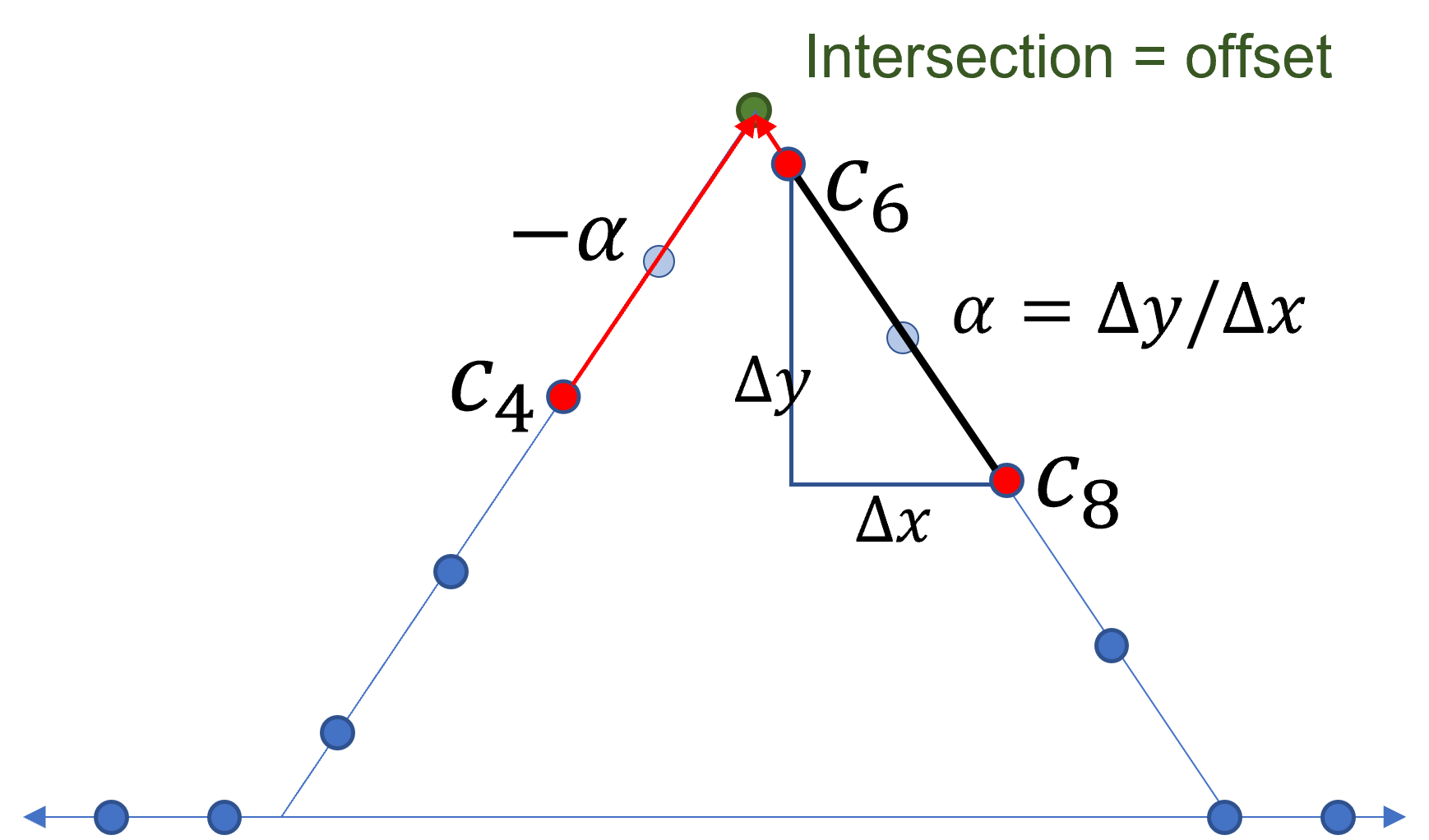}
      \caption{Illustrating the s-curve root-finding using linear
            interpolation.}
      \label{fig:s_curve_finding}
\end{figure}

To take advantage of the punctual correlator value as well as the early and late, we apply a linear interpolation to find the root of the s-curve (see fig. \ref{fig:s_curve_finding}). The two correlators located on the same flank (right side, in the example in \FigRef{fig:s_curve_finding}) are used to determine the slope of the correlation function ($\alpha:=\Delta y/\Delta x$ in \FigRef{fig:s_curve_finding}). The root is then determined by the intersection of two line-segments, one going through the punctual correlator with slope $\alpha$ and the second one going through the higher ``non-punctual'' correlator with slope $-\alpha$. Let us denote the early, late, and punctual correlators by $c_e$, $c_l$ and $c_p$, respectively. Further, define $c_{max} := \max(c_e, c_l)$ and $c_{min} := \min(c_e, c_l)$. Then the offset relative to the location of the punctual correlator can be written as:

\begin{equation}
      \Delta\tau = \frac{c_{max} - c_{min}}{c_p - c_{min}}\Delta\tau_c
\end{equation}

where $\Delta\tau_c$ is the correlator spacing. The location of the s-curve maximum is then given by $\tau = \tau_p + \Delta\tau$, where $\tau$ is the estimate of the offset of the maximum of the correlation function, and $\tau_p$ is the location of the punctual correlator.

\subsection{Vestigial Signal Search -- Exhaustive Search in E6\label{vestigial-signal-search-exhaustive-search-in-e6}}

Looking for vestigial signals, or in other words, multiple correlation peaks around the tracked signal, has been studied as a measure against spoofing \cite{hegarty2019spoofing,ahmed2023spoofing}. In our case, since an attacker must expect that an ACAS enabled receiver will verify a signal in E1 by looking for the corresponding signal in E6, the attacker must ensure its presence in E6 corresponding to the inauthentic signal in E1\footnote{If the user cannot verify the presence of the signal in E6, it will be declared as inauthentic.}. In that case two signals are present in E6: the true signal and the inauthentic signal. From a defense point of view, this means that we can expect to find both the true and the inauthentic signal in E6. Further, due to assumption 2 in section \ref{assumptions-on-the-receiver-architecture}, we also know that the \emph{earliest} signal within the time uncertainty interval has to be the true one.

Due to the delayed key disclosure concept of OSNMA\footnote{Remember that ACAS uses the OSNMA key to decrypt the RECS}, there is a time synchronization requirement on the receiver\cite{FernandezHernandez2023}. For one I/NAV sub-frame key delay, this is 30 seconds, which defines the search space for the VSS. Searching for the E6 signal over the entire initial time uncertainty (e.g., 30 seconds), for all possible code offsets, and over all possible Doppler bins, we refer to as an exhaustive search. An exhaustive search for vestigial signals in E6 is expensive in terms of computational load.

A further drawback is that due to the very large number of single-event hypotheses tests required, a threshold corresponding to a very low single-event false alarm probability must be used, in order to achieve an acceptable over-all false alarm probability. The effect of ACAS signal detection degradation due to high E6 search space was already discussed in \cite{terris2022operating}.

As an example, consider time uncertainty of $\Delta T=$ 30 seconds, a coherent integration time of $T_I =$ 4 ms, and the chip length $T_c =$ 60 m (corresponding to a BPSK(5) signal). The width of the frequency bin is then $\Delta F = 1/(2T_I) =$ 125 Hz \cite{borre2022gnss}.

If we wish to search every half chip, approximately $2\Delta T/T_c = 3\cdot 10^8$ chip offsets must be searched for each frequency bin. A Doppler uncertainty of $\pm$ 5 kHz, implies 80 Doppler bins to be searched. This yields a staggering number of $N=2.4 \cdot 10^{10}$ hypothesis to be calculated and tested.

As explained in \cite{FernandezHernandez2023}, the probability of false alarm corresponds to the case that an inauthentic signal is declared present when it is not. If we denote the single-event false alarm probability by $P_{\fa}$, then the overall false alarm probability $P_{\FA}$ is given by

\begin{equation}
      P_{\FA} = 1-(1-P_{\fa})^N \Rightarrow P_{\fa} = 1-(1-P_{\FA})^{1/N}
\end{equation}

For example, for the humble value of $P_{\FA}=10^{-3}$ for the overall false alarm probability, we obtain a required single-event false alarm probability of $P_{\fa}=4.16\cdot 10^{-14}$.

Theoretically, setting the detection threshold to this single-event false alarm, would generate a probability of detection of 90\% at a $C/N_0 =$ 40.12 dBHz for this example (see e.g.~\cite{Kaplan2006}).

Increasing the integration times to 8 ms or 16 ms would imply a doubling or quadrupling of the number of hypothesis, due to the narrower Doppler bins. Taking this into account, the required $C/N_0$ reduces to 37.2 dBHz and 34.27 dBHz, respectively, an improvement by about 2.9 dB in each step.

While those are quite reasonable values of $C/N_0$, it should be noted that they rely heavily on the strict Gaussian behavior of the noise process, since the threshold is set for \emph{very} low values of $P_{\fa}$. If for some reason the signal contains anything other than Gaussian noise, e.g., interference or multipath, then the behavior may change.

We therefore seek to develop an algorithm to significantly reduce the number of correlations to be evaluated, where advantage is taken of the assumption that any authentic signal in E6 must have a corresponding signal in E1.

\subsection{Vestigial Signal Search -- Handover from E1\label{sec:VSS-handover}}

The procedure is defined as a stand-alone process that relies only on samples from E1 and E6, and the assumption that the \emph{true} signal is detectable in E1.  After the results from the VSS are available, the tracking status of E1 can be updated.

\begin{figure}[!t]
      \centering
      \includegraphics[width=8cm]{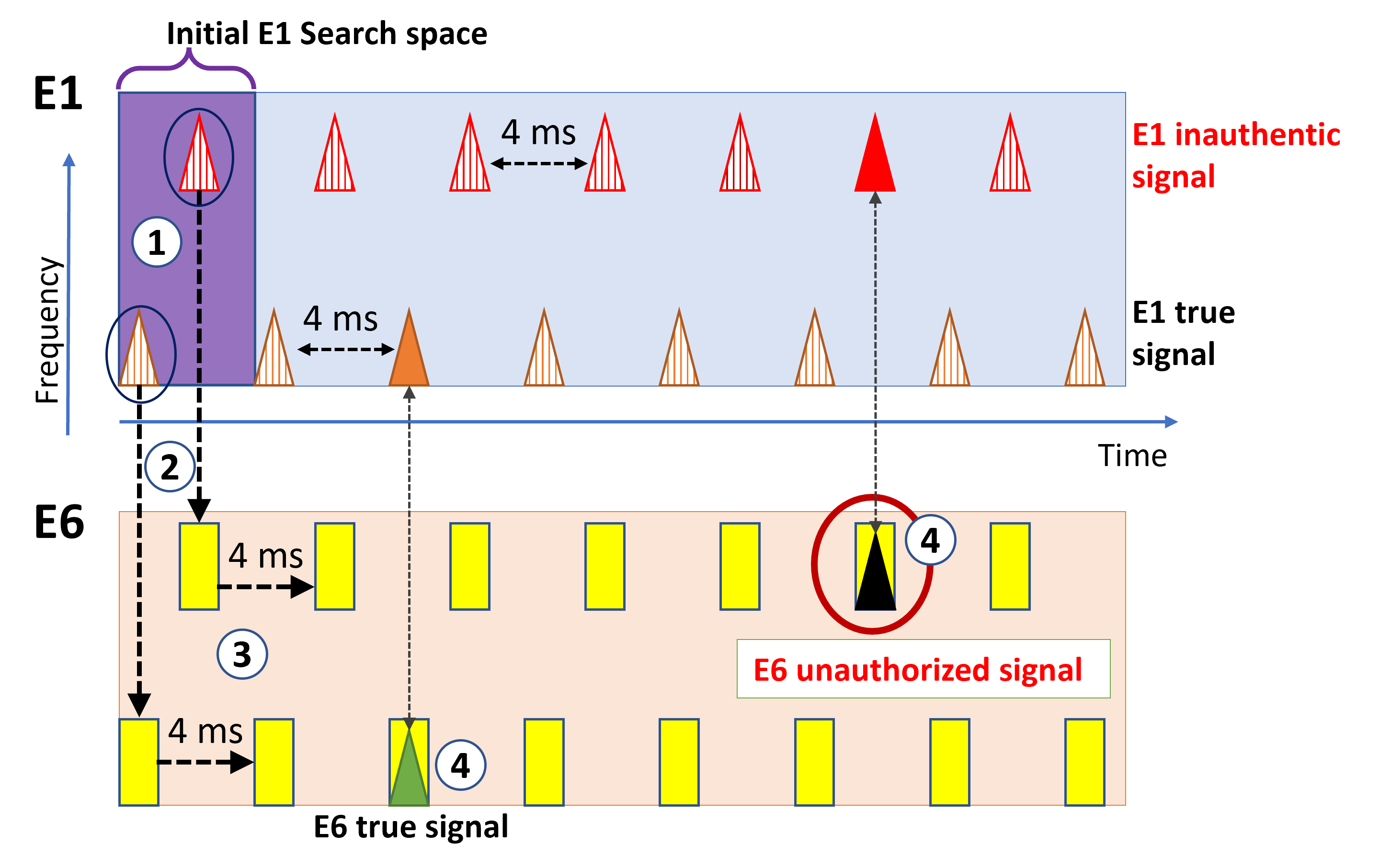}
      \caption{Illustrating the vestigial signal search (VSS) on E6 with
            hand-over from E1.}
      \label{fig:vss_handover}
\end{figure}

The vestigial signal search (VSS) concept is shown in \FigRef{fig:vss_handover}. The light blue area in the upper part of the figure symbolizes the E1 signal. It is assumed that a spoofing signal is present in both E1 and E6. The triangles in the upper part symbolize the 4 ms periodicity of the E1 signal and the solid colored ones represent the correct code phase ambiguity. The upper and lower row of triangles represent the inauthentic and true signals, respectively.

The idea is now to perform a full search in both time and frequency for one period of the E1 signal. Due to the periodicity of the E1 signal, we search over one period of E1, corresponding to the purple area in \FigRef{fig:vss_handover}. Normally, we expect to find only the true signal. However, if an inauthentic signal is present as well, two signals should be found: the true and the inauthentic one (two encircled triangles at 1 in \FigRef{fig:vss_handover}).

For any signal that is detected in E1, a hand-over is done to E6 (2 in \FigRef{fig:vss_handover}).

The E6 signal is aligned to the E1 signal, and must therefore be located at the same offset in E6, modulo 4 ms. Potential locations in E6 shown as yellow boxes in the lower part of \FigRef{fig:vss_handover}. After the handover, the search is then continued in E6, by incrementing the code offset by 4 ms (3 in \FigRef{fig:vss_handover}).

If more than one signal is found in E6 (4 in \FigRef{fig:vss_handover}), the receiver must compare the timing of the signals in E6 and select the earliest one and declare spoofing. Note that under the assumptions made, it is assumed safe to use the true signal, even in the presence of the spoofing signal.

As an alternative to the acquisition step in E1 (1 in \FigRef{fig:vss_handover}), the receiver could simply fully track both the true and the inauthentic E1 signal, resolve the code phase ambiguity (jumping straight to the solid triangles in the upper part of \FigRef{fig:vss_handover}.), and then do the hand-over, thus omitting the search over all the yellow boxes. This may be a less resource demanding alternative in some situations, but it would require a full extra channel and it would be more intrusive in an existing receiver design. Besides, we expect the presence of an inauthentic signal to be a seldom event, in which case the procedure terminates before the actual the handover to E6 is done. We mention both possibilities here as design choices.

Finally, it should be mentioned that the large search space (in the order of 30 seconds) must only be searched in the beginning, \emph{if} the search in E1 detects more than one signal. After the receiver has calculated the first authentic PVT, the time uncertainty is much reduced for subsequent searches.

Denoting, as before, the overall time uncertainty by $\Delta T= 30$s and the period of the E1 signal by $\delta T=10^{-3}$s, we can then estimate the number of hypothesis as follows $\Delta T / \delta T = 30/4\cdot 10^{-3} = 7500$ hypothesis per Doppler bin. Since we can predict the Doppler bin, there is only one Doppler bin to search. Thus, the hand-over search is about a million times less complex than the full exhaustive search\footnote{Using FFT based techniques, having the complexity $n\log(n)$ will dramatically reduce this ratio.}.

Comparing this to the case above with an overall false alarm probability of $P_{\FA}=10^{-3}$, we would require a single-event false alarm probability of $P_{\fa} = 1.33\cdot 10^{-7}$ (compared to $P_{\fa}=4.16\cdot 10^{-14}$ for the exhaustive search case), which would lead to a required $C/N_0=37.66$ dBHz for the same overall false alarm probability (compared to $C/N_0=40.12$ dBHz for the exhaustive search case.). The required $C/N_0$ is then about 2.46 dB less than in the exhaustive search case.

However, if the attacker were able to mask out or null the true signal on E1, this would potentially cause the victim receiver to miss the true signal and only discover the spoofing signal during the handover process, or stay locked to a meaconing E6 signal.

\section{ACAS Threat Mitigation Levels\label{sec:ACAS_mitigation}}

We define a set of measures that would give a good level of protection against as large a set of attacks as possible. Since applications may have different levels of required protection, we introduce three mitigation levels. Level one is limited to the direct processing of the data (RECS, BGD, OSNMA) provided by ACAS. The second level includes more defenses to mitigate more threats, at reasonable cost and effort, and at the third mitigation level, we try to maximize the defenses.

\subsection{Mitigation Level 1\label{mitigation-level-1}}

We first consider a ``minimum defense strategy'' that we refer to as a ``level 1 defense''. This procedure is limited to only check for the ACAS signal and compare it to the E1 measurements:

\begin{enumerate}
      \item Check that significant E6 correlation power is obtained at the expected offset, based on the E1 measurement using the algorithms in \SecRef{maximum-signal-detection} or \SecRef{early-signal-detection}. The test should be against a threshold established from the post-correlation noise output. If the check is against the correlator output using a code not present in the signal, this will also defend against attacks that simply push the correlation power up\footnote{There may also be an issue with the RF chain approaching saturation.}.
      \item Check that the difference in pseudorange between E1 and E6 is within an acceptable range. This can be done in snapshot mode on E6 using the algorithm described in \SecRef{sec:root_finding}. The resulting offset relative to the corresponding pseudorange measurement on E1 is compared against a threshold, as outlined in \cite{FernandezHernandez2023}.
\end{enumerate}

This will defend against any spoofing attack that operates on E1 only, since the check on E6 would raise an alarm with the corresponding probabilities ($P_{\FA}$ and $P_d$) in this case.

We expect any attack that manages to place the E6 encrypted signal consistently with the E1 measurements to be successful against the level 1 strategy. A simple meaconing attack on both frequencies could achieve this.

\subsection{Mitigation Level 2\label{mitigation-level-2}}

The next level aims to further reduce the set of successful attacks. The following steps are assumed to be performed in \emph{addition} to the previously described steps:

\begin{enumerate}
      \setcounter{enumi}{2}
      \item Perform AGC and $C/N_0$ monitoring. See e.g. discussion in \cite{Lo2021}.
      \item Look for vestigial signals with handover from E1 (see sec.~\ref{sec:VSS-handover})
      \item PVT-level monitoring
            \begin{itemize}
                  \item Monitor the clock drift
                  \item RAIM check
            \end{itemize}
\end{enumerate}

The level 2 strategy is expected to detect a plain jamming attack, due to item 3 above.

When a RECS is available, we expect this strategy to defend against a large set of attacks. However, it is possible that a very advanced attacker uses a smart jammer to null out the true signal in E1. Therefore, the assumption on the simultaneous presence of signals in E1 and E6 may \emph{not always} hold, which is the basis for the VSS with handover from E1.

Also, all the measurements between RECS are not protected, which would require continuous monitoring. The AGC \& $C/N_0$ monitoring and the PVT-level monitoring will provide some protection between RECS, but we consider the metrics proposed in \cite{Wesson2018} even better suited for this.

As explained in \cite{Curran2017}, a well-tuned weak jammer could prevent the user from accessing the OSNMA data on I/NAV, by jamming the signal just enough to generate enough bit errors, such that the data cannot be correctly demodulated. We expect such an unnoticed denial-of-service attack to be successful under the level 2 strategy.

\subsection{Mitigation Level 3\label{mitigation-level-3}}

With the third level strategy we try to remove any residual threat. We suggest the following enhancement of the level 2 mitigation strategy, where steps 4 and 5 replace the corresponding steps in mitigation level 2.

\begin{enumerate}
      \setcounter{enumi}{3}
      \item Obtain assisted OSNMA (ANMA)\cite{ODriscoll2023}\footnote{Although ANMA does not exist as an official service, we reference it here as it has been analyzed in the literature, and we consider such a service to be very useful.}. Although the ACAS concept is designed for unconnected users, this step is still considered valid in the ACAS context since:
            \begin{itemize}
                  \item The data rate implied for ANMA is much lower than downloading the RECS.
                  \item An on-line version of ACAS is also conceivable for connected users.
            \end{itemize}
      \item Look for vestigial signals using an exhaustive search in E6 (see sec.~\ref{vestigial-signal-search-exhaustive-search-in-e6}), to counter the possible nulling of the true signal in E1.
      \item PVT-level monitoring:
            \begin{itemize}
                  \item Monitor the Clock Drift
                  \item RAIM Check
            \end{itemize}
      \item Bridging between ACAS authentications. To defend against this, we propose to add the detection metrics researched in \cite{Wesson2018}. Using additional sensors like IMU can also be helpful here.
\end{enumerate}

Due to item 4, the OSNMA bit error attack discussed above \cite{Curran2017} could be detected, but not mitigated.

\subsection{Summary and Residual Threats\label{summary-and-residual-threats}}

\begin{figure}[!t]
      \centering
      \includegraphics[width=85mm]{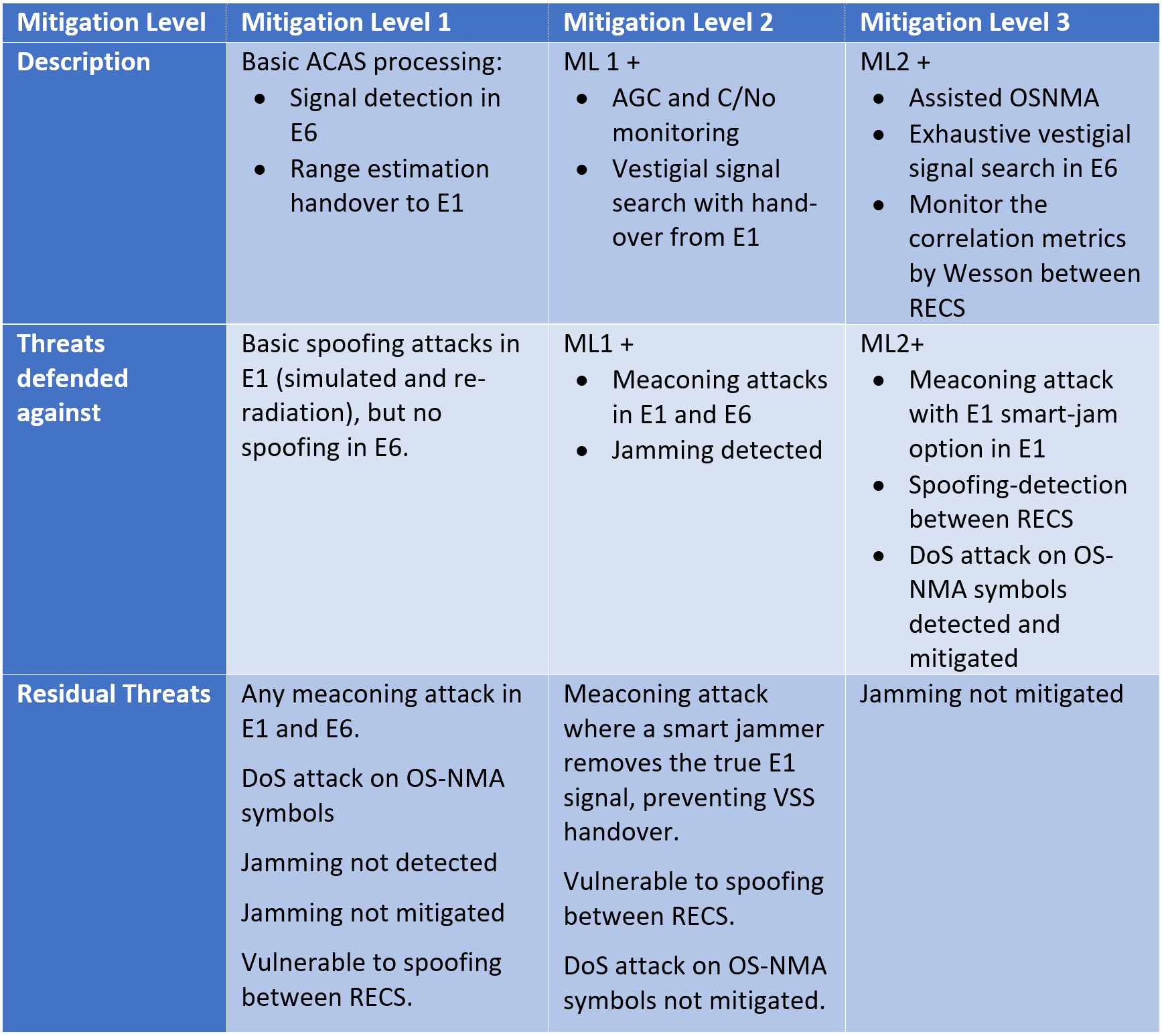}
      \caption{Overview of mitigation levels, threats mitigated and the
            residual threats.}
      \label{fig:mit_levels}
\end{figure}

In \FigRef{fig:mit_levels} the three mitigation levels are summarized, together with threats that the mitigation levels are expected to defend against and residual threats. Note that it is thinkable that there exist threats that have not been taken into account. It is also not guaranteed that the threats are mitigated to 100\%, as there will always be a residual risk that a particular defense is penetrated. This is related to the $P_{\fa}$ and $P_d$ previously discussed.

Regarding the residual threat of level 3, it is always assumed that given enough interference power, it will always be possible to create enough interference such that a PVT cannot be measured, regardless of the defense.

\section{ACAS Modelling\label{sec:ACAS_Modelling}}

To verify the user algorithms and the mitigation levels, we developed a simulator with the following main objectives:

\begin{itemize}
      \item Analyze the performance of the ACAS Authentication.
      \item Emulate the receiver implementation concerning algorithms involving the measurements.
      \item Provide a framework to implement simulated spoofing attacks.
\end{itemize}

\begin{figure}[!t]
      \centering
      \includegraphics[width=8cm]{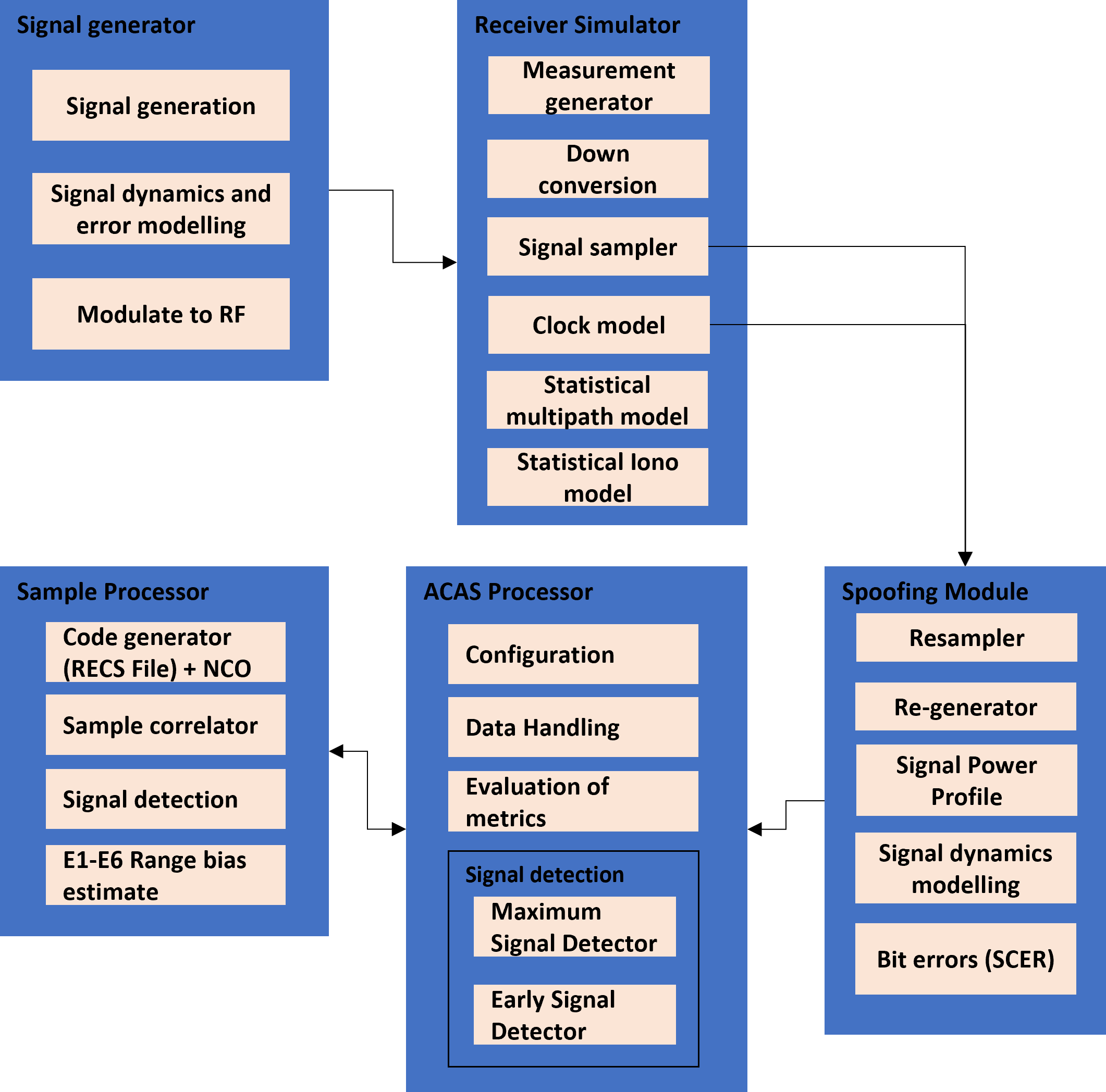}
      \caption{Overview of the ACAS Simulator
            architecture.\label{fig:sim_arch}}
\end{figure}

Note that the simulator is fundamentally a single line-of-sight simulation, and our results are complemented by real-live data from e.g. \cite{TerrisGallego2023}.
In \FigRef{fig:sim_arch} an overview of the ACAS simulator is given. The main components are:

\begin{itemize}
      \item Signal generator: Generates signal samples and applies range modelling
      \item Receiver simulation: Emulates the sampling process, applies error modelling (multipath, ionosphere etc.)
      \item Spoofing module: Models spoofing signals and adds them to the ``true'' signal.
      \item ACAS processor: High-level steering of the ACAS processing. Uses the sample processor for low-level signal processing
      \item Sample processor: Used by the ACAS processor to perform signal processing (mainly RECS correlation) collected by the receiver model.
\end{itemize}

The signal generator module modulates the codes onto a complex RF signal. The modulator is implemented as a continuous complex function of time that can be sampled with arbitrary sample rates. Doppler is applied on both code and carrier.

The simple line-of-sight range model in Eq.\ref{eq:dyn} is applied to the samples. This model is well suited to generate all kinds of dynamic situations as well as any corner case, including user dynamics. However, this model does not accurately model any orbit or user dynamics.

The ACAS Processor receives observables and the sample batches from the receiver model and then performs the signal detection in E6 (see \SecRef{sec:ACAS_user_algorithms}).

The receiver model generates pseudorange, carrier phase and Doppler measurements on E1 and samples on E6, with the input from the spoofing module. The following error models are applied in the simulation:

\begin{itemize}
      \item Two-state receiver clock model, based on the Allan variance parameters $h_0$ and $h_{-2}$. In all simulations, the following values were used\cite{McNeill2017}: $h_0 = 7.115\cdot 10^{-24}$, $h_{-2} = 4.311\cdot 10^{-21}$.
      \item Statistical multipath model described in \cite{Wesson2018}. This model was derived from the Land Mobile Satellite Channel Model (LMSCM) described in \cite{Steingass2004,Lehner2007}. The model is applied on sample-level to the simulated E6 samples.
      \item Statistical ionosphere error model, based on measurement campaigns performed by the gAGE group \cite{Sanz2017,RoviraGarcia2019}. The dataset spans a network of over 180 reference stations world-wide, and the two full data sets of the years 2014 (solar maximum) and 2017 (average solar activity) were used.
      \item Noise modelling on the E1 measurements, based on the Betz theory \cite{J.W.Betz2000c}.
      \item Gaussian white noise is added to the samples, based on the configured $C/N_0$ and bandwidth.
\end{itemize}

\section{Simulation Results\label{simulation-results}}

Out of the many simulations performed, in this section we present the results of two simulations, which we consider the most representative at this stage: A non-spoofing scenario, and an advanced spoofing scenario with a zero-delay attack, followed by signal lift-off. Both include all error models and harsh signal conditions ($C/N_0 = $ 35 dBHz). For the line-of-sight dynamics the following model is used:
\begin{equation}
      s(t) = s_0 + v t + a t^2/2 + A\sin(\omega t)
      \label{eq:dyn}
\end{equation}

with the following parameters:

\begin{itemize}
      \item $s_0$ = 0 m
      \item $v$ = 51.023 m/s
      \item $a$ = 0 m/s\textsuperscript{2}
      \item $A$ = 0.5012 m
      \item $\omega = 2\pi\cdot 0.5012$ rad/s
\end{itemize}

The parameters above imply a maximum line-of-sight acceleration of about $9g$. The model could be considered quite dynamic and worse than nominal, but it is maintained here for a more worst-case consideration.

\subsection{ACAS Performance under Harsh Conditions, no Spoofing\label{acas-performance-under-harsh-conditions}}

The detection threshold for the single event detection for the RECS is set to $P_{\fa}=10^{-7}$. The ionospheric model uses the statistical data from 2014 (high solar activity)\cite{RoviraGarcia2015,RoviraGarcia2019}. The multipath model was set to the range of elevations of [5..30] degrees. The simulation was run for 100,000 seconds, using a RECS length of 16 ms, which corresponds to the maximum RECS length currently expected in the specification \cite{FernandezHernandez2023}. The number of correlators was set to 11 with a correlator spacing of 15 m, about a quarter of an E6-C BPSK(5) chip.

\begin{figure}[!t]
      \centering
      \includegraphics[width=8cm]{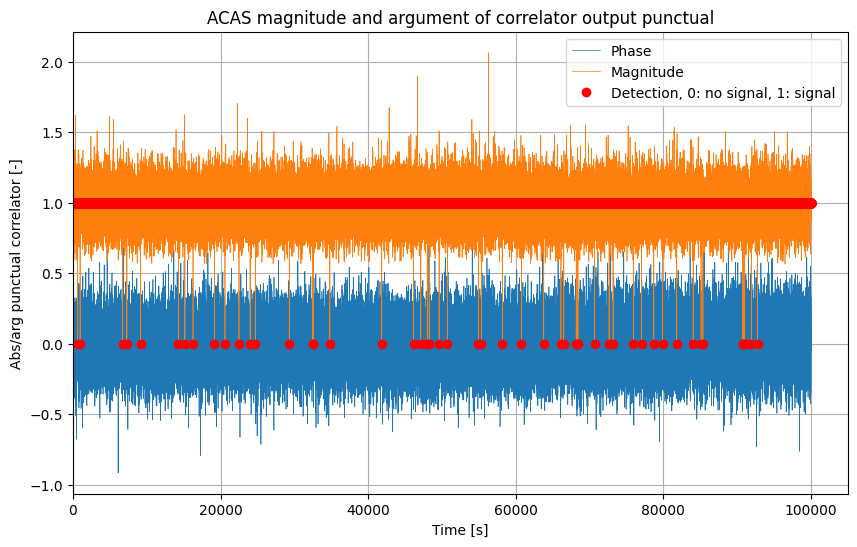}
      \caption{Output of the detecting correlator. The orange curve:
            magnitude, blue curve: phase. The red dots indicate signal detection (1:
            Signal detected, 0: Signal not detected.\label{fig:corr_harsh}}
\end{figure}

Figure \ref{fig:corr_harsh} shows the output of the correlator that was selected to detect the signal.
The red dots indicate the status of the signal detection: 1: signal was detected, 0: signal was not detected. 52 missed detection events were recorded for the 100,000 measurements, indicating a detection probability of $P_d =$ 99.95\%. The theoretical probability of detection for the corresponding noise-only case (i.e., $C/N_0=$ 35 dBHz, $T_I=10^{-7}$, $P_{\fa}=10^{-7}$) is $P_d=$ 99.9995\% and that for $C/N_0=$ 34 dBHz is $P_d=$ 99.96\%.

\begin{figure}[!t]
      \centering
      \includegraphics[width=8cm]{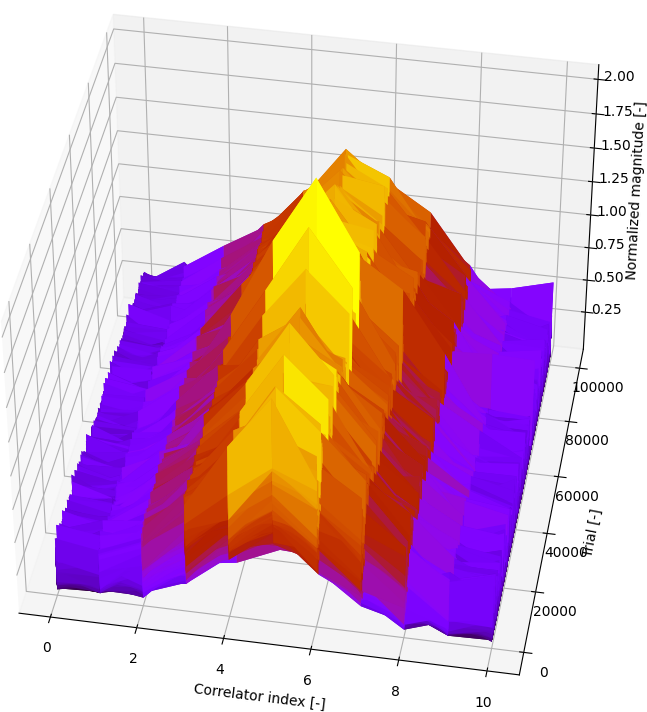}
      \caption{Waterfall plot of the magnitude of the correlation function. 11
            correlators were used with a correlator spacing of 15
            m.\label{fig:ccf_harsh}}
\end{figure}

\begin{figure}[!t]
      \centering
      \includegraphics[width=8cm]{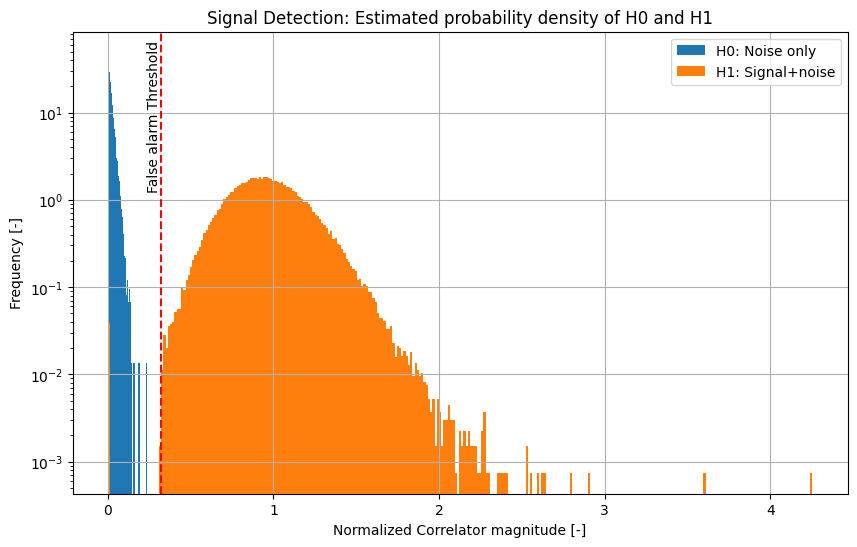}
      \caption{Histogram - Signal detection for the nominal scenario under
            harsh signal conditions ($C/N_0=$ 35 dBHz).\label{fig:hist_harsh}}
\end{figure}

Figures \ref{fig:ccf_harsh} and \ref{fig:hist_harsh} show the correlator output and a histogram of the magnitude of the detecting correlator, respectively. It should be noted that the histogram for the signal detection events is cut off at the detection threshold. This is because the signal is not defined if none of the correlators are above the threshold.

\begin{figure}[!t]
      \centering
      \includegraphics[width=8cm]{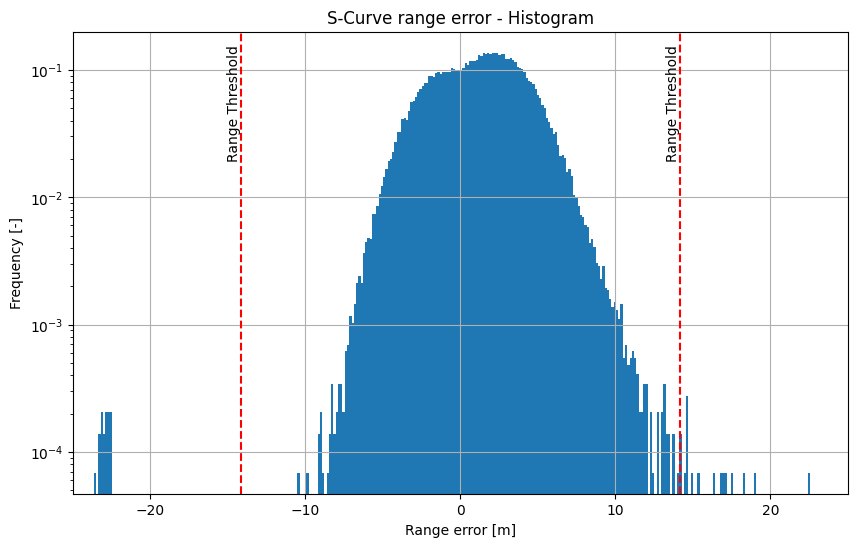}
      \caption{Histogram - Range detection for the nominal scenario under
            harsh signal conditions ($C/N_0=$ 35
            dBHz).\label{fig:hist_range_harsh}}
\end{figure}

\begin{figure}[!t]
      \centering
      \includegraphics[width=8cm]{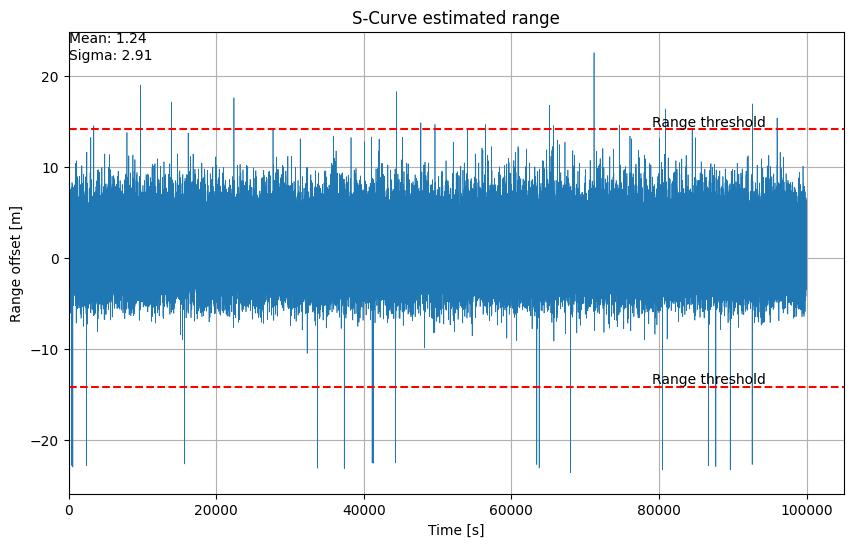}
      \caption{Time series of the range detection for the nominal scenario
            under harsh signal conditions ($C/N_0=$ 35
            dBHz).\label{fig:time_series_harsh}}
\end{figure}

\FigRef{fig:hist_range_harsh} and \FigRef{fig:time_series_harsh} show the s-curve range detection. \FigRef{fig:hist_range_harsh} shows a histogram in logarithmic scale and \FigRef{fig:time_series_harsh} shows the same data as a time-series. The effect of the ionospheric error model is seen on the right side. The right flank of the histogram falls off linearly. This is an indication that the distribution of the ionospheric error is heavy tailed, invalidating Gaussian approximations (see e.g. \cite{Larson2018} on Gaussian over-bounding of heavy-tailed distributions).

To the left on the histogram there is a very small accumulation around -22.5 m. These events correspond to events where the early acquisition algorithm triggers on the left flank of the correlation function.

\subsection{ACAS Performance under Harsh Conditions and Spoofing\label{acas-performance-under-spoofing-conditions}}

This section presents a simulated spoofing attack to demonstrate the feasibility of the ACAS processing under such conditions. We consider the same environment as in the previous simulation but including the spoofer. Two separate runs are performed: one using the maximum signal detection algorithm and one using the early signal detection algorithm.

\begin{figure}[!t]
      \centering
      \includegraphics[width=8cm]{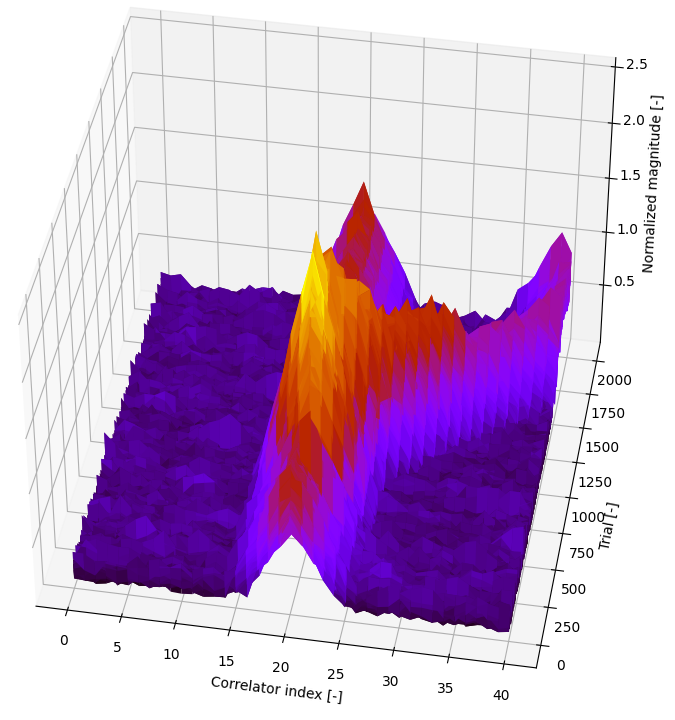}
      \caption{Waterfall plot of the correlators for the spoofing attack
            scenario under harsh conditions ($C/N_0=$ 35
            dBHz).\label{fig:ccf_spoof}}
\end{figure}

The spoofing scenario is shown in the figure \ref{fig:ccf_spoof}. The profile of the spoofing signal starts at the front of the figure, and the distance between the true signal and the spoofing signal is modelled as a sine wave, approaching the true signal from the right.

The power of the spoofing signal is very low, both when approaching the true signal and when it is directly on top of the true signal. This makes sense, as it prevents excessive fading and too high amplitude. Before the spoofing signal ``leaves'' the true signal, the magnitude must be increased relatively fast to sustain a self-standing signal when both signals are not (or only partly) added.

Note that this kind of spoofing attack would be \emph{very difficult} to mount, and depending on the scenario, in some cases even impossible, as the true signal will always arrive before the spoofing signal (assumption 2 above), while for a zero-delay attack
the delay of the inauthentic signal can be made arbitrarily small. Note that just one E6-C chip of delay ($0.195\mu$s, or 58.7m) may render this attack impossible.

\begin{figure}[!t]
      \centering
      \includegraphics[width=8cm]{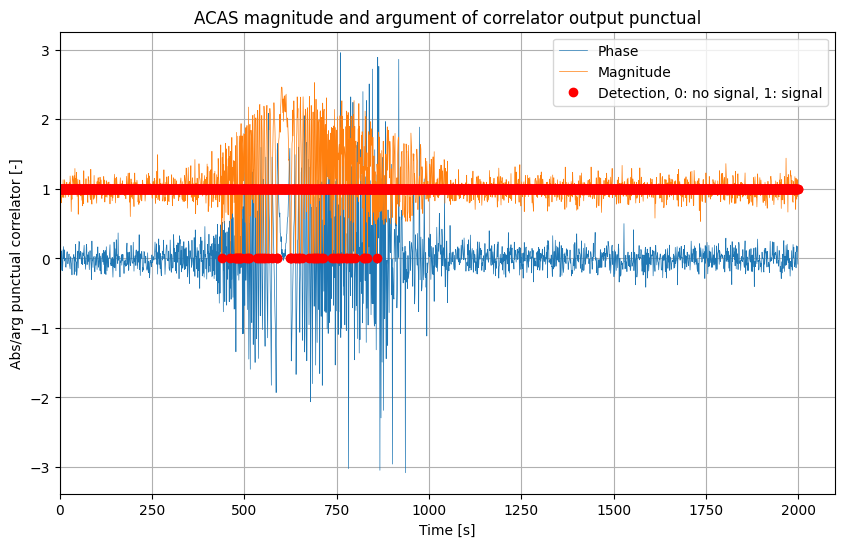}
      \caption{Early detector - Output of the detecting correlator for the
            spoofing attack scenario.\label{fig:corr_early_spoof}}
\end{figure}

\begin{figure}[!t]
      \centering
      \includegraphics[width=8cm]{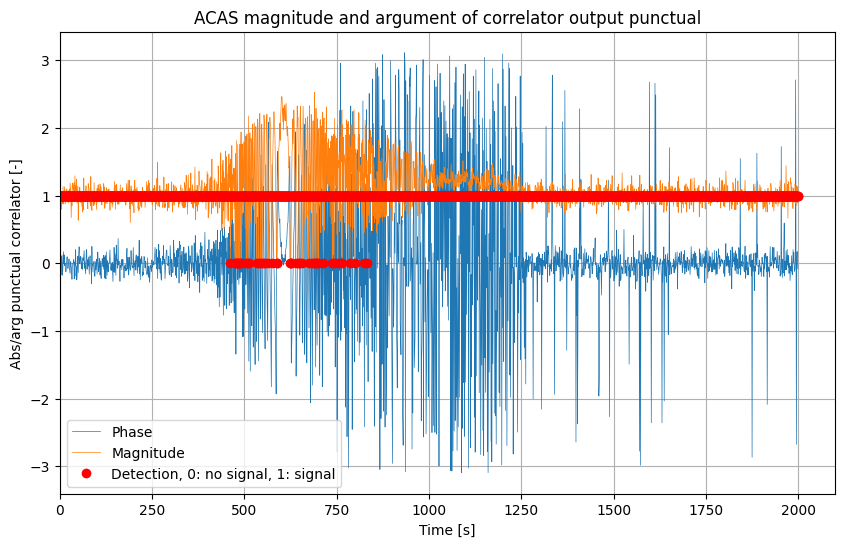}
      \caption{Maximum detector - Output of the detecting correlator for the
            spoofing attack scenario.\label{fig:corr_max_spoof}}
\end{figure}

The plots in figures \ref{fig:corr_early_spoof} and \ref{fig:corr_max_spoof} show the output of the detecting correlator, its phase and the signal detection status. In figure \ref{fig:corr_early_spoof} the results from using the early detector algorithm are shown, and in figure \ref{fig:corr_max_spoof} the result from the maximum detector can be seen. In both figures, the transition region at $T=$ 600 seconds is evident, in which the spoofer attempts to carry off the signal. As the spoofer comes in, the phase starts rotating and at the spot where the spoofer transitions from approaching to pulling away, there is a small region where the phase and the changes in amplitude calm down a bit. In case of the maximum detector, the phase continues to rotate after the spoofer starts to move away.

\begin{figure}[!t]
      \centering
      \includegraphics[width=8cm]{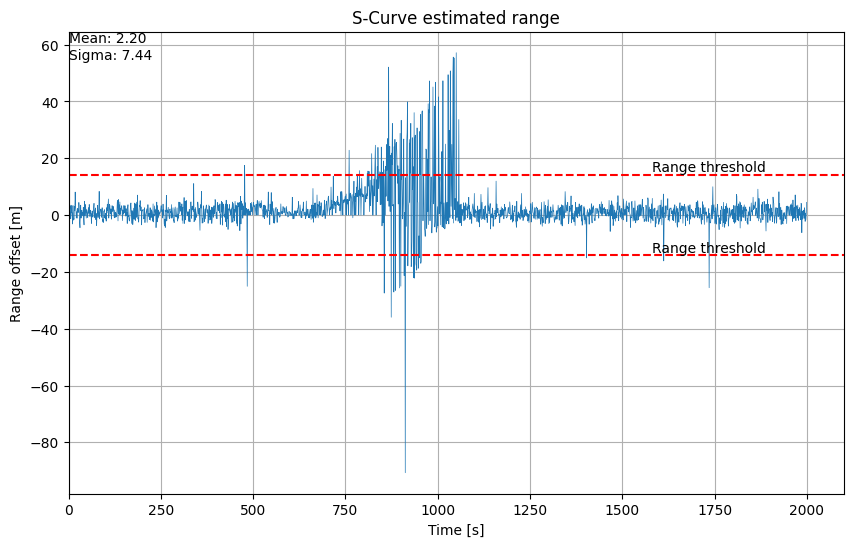}
      \caption{Early detector - Estimate of the s-curve offset for the
            spoofing attack scenario.}
      \label{fig:s_curve_early}
\end{figure}

\begin{figure}[!t]
      \centering
      \includegraphics[width=8cm]{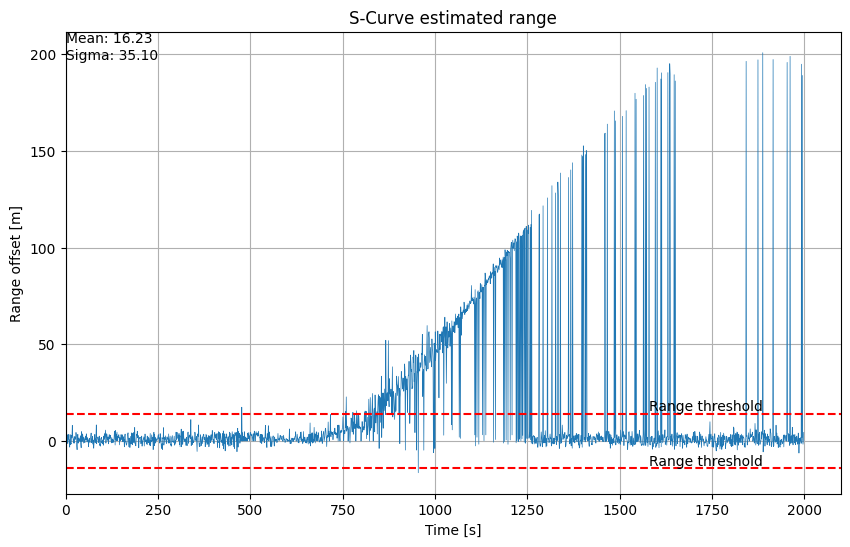}
      \caption{Maximum detector - Estimate of the s-curve offset for the
            spoofing attack scenario.}
      \label{fig:s_curve_max}
\end{figure}

In \FigRef{fig:s_curve_early} and \FigRef{fig:s_curve_max}, the estimate of the s-curve error (using the true E1 signal as a reference) is shown. At the beginning when the spoofer sets in, between $T=$ 500 seconds and $T=$ 750 seconds, we notice a small upward curve in the range estimate in both cases.

Then when the spoofer turns up the power, it captures the range estimate at the beginning in both cases. However, in the case of the early detector, the range estimate goes back to the baseline after about $T=$ 1050 seconds, after some fluctuations. For the maximum detector, this is not the case and it follows the spoofer quite well. Note the toggling between the spoofer and the baseline. There are at least three reasons for this:

\begin{itemize}
      \item Fading between the true and the spoofing signal
      \item Thermal noise: sometimes the noise is higher in the ``true correlator'' and sometimes it is higher in the ``spoofing correlator''
      \item The spacing between correlators: When the maximum of the signals are located on a correlator, the magnitude is maximized.
\end{itemize}

Note the one outlier in \FigRef{fig:s_curve_early} at around $T=800$ seconds. This is the same effect observed in \FigRef{fig:hist_range_harsh} and \FigRef{fig:time_series_harsh}, where the signal is detected on the flank of the correlation function.

In summary, in the case of spoofing, even a zero-delay E6-meaconing attack can be detected from the beginning, as shown in \FigRef{fig:corr_early_spoof}, and mitigated after some minutes, as shown in \FigRef{fig:s_curve_early}.

\section{Conclusions\label{conclusions}}
The purpose of this work was to assess to which extent Galileo ACAS can protect against spoofing, and how it can be strengthened by additional receiver checks.

The receiver can assume that the true E6-C encrypted signal always arrives before any other signal. Considering this assumption, two user algorithms are presented and analyzed. Firstly, the proposed \textit{early signal detection} algorithm applied on the E6-C snapshot correlators detects the true E6-C signal even in the presence of a stronger later spoofing signal. Secondly, \textit{vestigial signal search}  (VSS) strategies are proposed and optimized. Vestigial search is applied first in E1, before handing over to E6-C. The periodicity of the E1 open code reduces the computational complexity be a factor of a million, and increases the sensitivity by about 2.5 dB compared to an exhaustive search in E6.

We then define three receiver mitigation levels. The first one is limited to basic ACAS processing, simply checking the presence of the E6 encrypted signal. The second level adds E1-based vestigial signal search, $C/N_0$ and total noise power monitoring, and E6-C early signal detection. Residual risks (e.g. masking out E1 or other denial of service attacks) are covered by the third level.

A detailed line-of-sight simulator was implemented to test the performance of the user algorithms. The simulator models ionosphere, multipath, noise on both measurements and samples, receiver clock and user dynamics. An advanced spoofing attack, corresponding to a signal lift-off attack, was implemented as well. We present results of two tests, both under harsh conditions, including a $C/N_0=$ 35 dBHz, multipath and high ionospheric activity. The second simulation introduces a lift-off type of spoofer under the same conditions.

The simulation results show that, in the absence of spoofing, ACAS RECS performs reliably even in harsh conditions and with realistic ionosphere, multipath and dynamic models. In particular, with 16-ms RECS and a false alarm probability of $10^{-7}$, a probability of detection of $P_d =$ 99.95\% is obtained. In the case of spoofing, and under similar conditions, a zero-delay E6 meaconing attack was detected from the beginning and mitigated after some minutes, allowing the receiver to continue operation even under spoofing conditions.

Further work will include expanding VSS and early signal detection testing with further configurations and scenarios, and  testing ACAS and the proposed methods with real signals.

\section{Disclaimer\label{disclaimer}}
The content of this article does not necessarily reflect the official position the authors’ organizations. Responsibility for the information and views set out in this article lies entirely with the authors. In particular, future ACAS publications, reference algorithms, standards or user scenarios may differ from the algorithms or techniques proposed in this work.

\bibliographystyle{IEEEtran}
\bibliography{IEEEabrv,acas_paper}

\begin{thebibliography}{10}
\providecommand{\url}[1]{#1}
\csname url@samestyle\endcsname
\providecommand{\newblock}{\relax}
\providecommand{\bibinfo}[2]{#2}
\providecommand{\BIBentrySTDinterwordspacing}{\spaceskip=0pt\relax}
\providecommand{\BIBentryALTinterwordstretchfactor}{4}
\providecommand{\BIBentryALTinterwordspacing}{\spaceskip=\fontdimen2\font plus
\BIBentryALTinterwordstretchfactor\fontdimen3\font minus \fontdimen4\font\relax}
\providecommand{\BIBforeignlanguage}[2]{{%
\expandafter\ifx\csname l@#1\endcsname\relax
\typeout{** WARNING: IEEEtran.bst: No hyphenation pattern has been}%
\typeout{** loaded for the language `#1'. Using the pattern for}%
\typeout{** the default language instead.}%
\else
\language=\csname l@#1\endcsname
\fi
#2}}
\providecommand{\BIBdecl}{\relax}
\BIBdecl

\bibitem{fernandez2016navigation}
I.~Fern{\'a}ndez-Hern{\'a}ndez, V.~Rijmen, G.~Seco-Granados, J.~Simon, I.~Rodr{\'\i}guez, and J.~D. Calle, ``A navigation message authentication proposal for the galileo open service,'' \emph{NAVIGATION: Journal of the Institute of Navigation}, vol.~63, no.~1, pp. 85--102, 2016.

\bibitem{EUSPA_OSNMA}
``Tests of {Galileo} {OSNMA} underway,'' \url{http://https://www.euspa.europa.eu/newsroom/news/tests-galileo-osnma-underway}, accessed: 2021-02-11.

\bibitem{Anderson2017}
J.~M. Anderson, K.~L. Carroll, N.~P. DeVilbiss, J.~T. Gillis, J.~C. Hinks, B.~W. O'Hanlon, J.~J. Rushanan, L.~Scott, and R.~A. Yazdi, ``Chips-message robust authentication (chimera) for {GPS} civilian signals,'' in \emph{Proceedings of the 30th International Technical Meeting of The Satellite Division of the Institute of Navigation ({ION} {GNSS+} 2017)}.\hskip 1em plus 0.5em minus 0.4em\relax Institute of Navigation, nov 2017.

\bibitem{fernandez2021analysis}
I.~Fern{\'a}ndez-Hern{\'a}ndez, T.~Ashur, and V.~Rijmen, ``Analysis and recommendations for mac and key lengths in delayed disclosure gnss authentication protocols,'' \emph{IEEE Transactions on Aerospace and Electronic Systems}, vol.~57, no.~3, pp. 1827--1839, 2021.

\bibitem{scott2003anti}
L.~Scott, ``Anti-spoofing \& authenticated signal architectures for civil navigation systems,'' in \emph{Proceedings of the 16th International Technical Meeting of the Satellite Division of The Institute of Navigation (ION GPS/GNSS 2003)}, 2003, pp. 1543--1552.

\bibitem{Lo2009}
S.~Lo, D.~D. Lorenzo, P.~Enge, D.~Akos, and P.~Bradley, ``{Signal Authentication: A Secure Civil GNSS for Today},'' \emph{Inside GNSS}, Sep. 2009.

\bibitem{fernandez2022semi}
I.~Fernandez-Hernandez, S.~Cancela, R.~Terris-Gallego, G.~Seco-Granados, J.~A. L{\'o}pez-Salcedo, C.~O'Driscoll, J.~Winkel, A.~d. Chiara, C.~Sarto, V.~Rijmen \emph{et~al.}, ``Semi-assisted signal authentication based on galileo acas,'' \emph{arXiv preprint}, 2022.

\bibitem{terris2022operating}
R.~Terris-Gallego, J.~A. L{\'o}pez-Salcedo, G.~Seco-Granados, and I.~Fernandez-Hernandez, ``Operating modes and performance evaluation of galileo assisted commercial authentication service,'' in \emph{Proceedings of the 35th International Technical Meeting of the Satellite Division of The Institute of Navigation (ION GNSS+ 2022)}, 2022, pp. 3393--3407.

\bibitem{TerrisGallego2023}
R.~Terris-Gallego, J.~A. Lopez-Salcedo, G.~Seco-Granados, and I.~Fernandez-Hernandez, ``{E1-E6 SDR platform based on bladeRF for testing Galileo Assisted Commercial Authentication Service},'' in \emph{Proc. European Navigation Conference (ENC)}, Jun. 2023.

\bibitem{ardizzon2022authenticated}
F.~Ardizzon, L.~Crosara, N.~Laurenti, S.~Tomasin, and N.~Montini, ``Authenticated timing protocol based on galileo acas,'' \emph{Sensors}, vol.~22, no.~16, p. 6298, 2022.

\bibitem{FernandezHernandez2023}
I.~Fernandez-Hernandez, J.~Winkel, C.~O'Driscoll, S.~Cancela, R.~Terris-Gallego, J.~A. López-Salcedo, G.~Seco-Granados, A.~D. Chiara, C.~Sarto, D.~Blonski, and J.~D. Blas, ``Semi-assisted signal authentication for {Galileo}: Proof of concept and results,'' \emph{IEEE Transactions on Aerospace and Electronic Systems}, pp. 1--13, 2023.

\bibitem{J.W.Betz2000c}
J.~W. Betz and K.~R. Kolodziejski, ``{Extended Theory of Early-Late Code Tracking for Bandlimited GPS Receiver},'' \emph{Navigation: Journal of The Institute of Navigation}, vol.~47, no.~3, pp. 211--226, 2000.

\bibitem{RoviraGarcia2015}
A.~Rovira-Garcia, J.~M. Juan, J.~Sanz, G.~Gonz{\'{a}}lez-Casado, and D.~Ib{\'{a}}{\~{n}}ez, ``Accuracy of ionospheric models used in {GNSS} and {SBAS}: methodology and analysis,'' \emph{Journal of Geodesy}, vol.~90, no.~3, pp. 229--240, oct 2015.

\bibitem{RoviraGarcia2019}
\BIBentryALTinterwordspacing
A.~Rovira-Garcia, D.~Ibáñez-Segura, R.~Orús-Perez, J.~M. Juan, J.~Sanz, and G.~González-Casado, ``Assessing the quality of ionospheric models through {GNSS} positioning error: methodology and results,'' \emph{GPS Solutions}, vol.~24, no.~1, p.~4, 2019. [Online]. Available: \url{https://doi.org/10.1007/s10291-019-0918-z}
\BIBentrySTDinterwordspacing

\bibitem{Kaplan2006}
E.~D. Kaplan and C.~J. Hegarty, \emph{{Understanding GPS Principles and Applications}}.\hskip 1em plus 0.5em minus 0.4em\relax Artech House, 2006.

\bibitem{Psiaki2016}
M.~Psiaki and T.~Humphreys, ``{GNSS Spoofing and Detection},'' in \emph{Proceedings of the IEEE}, vol. 104, no.~6, 2016, pp. 1258--1270.

\bibitem{Risueno2005}
G.~Lopez-Risueno and G.~Seco-granados, ``Measurement and processing of indoor gps signals using one-shot software receiver,'' in \emph{Proceedings of ESA NAVITEC, Noordwijk, NL, pp. 1–9, 2004.}\hskip 1em plus 0.5em minus 0.4em\relax Noordwijk, The Netherlands: ESA, December 2004.

\bibitem{borre2022gnss}
K.~Borre, I.~Fernandez-Hernandez, J.~A. L{\'o}pez-Salcedo, and M.~Z.~H. Bhuiyan, \emph{GNSS Software Receivers}.\hskip 1em plus 0.5em minus 0.4em\relax Cambridge University Press, 2022.

\bibitem{hegarty2019spoofing}
C.~Hegarty, B.~O’Hanlon, A.~Odeh, K.~Shallberg, and J.~Flake, ``Spoofing detection in gnss receivers through cross-ambiguity function monitoring,'' in \emph{Proceedings of the 32nd International Technical Meeting of the Satellite Division of The Institute of Navigation (ION GNSS+ 2019)}, 2019, pp. 920--942.

\bibitem{ahmed2023spoofing}
S.~Ahmed, S.~Khanafseh, and B.~Pervan, ``Spoofing detection using decomposition of the complex cross ambiguity function with measurement correlation,'' in \emph{2023 IEEE/ION Position, Location and Navigation Symposium (PLANS)}.\hskip 1em plus 0.5em minus 0.4em\relax IEEE, 2023, pp. 500--510.

\bibitem{Lo2021}
S.~Lo, F.~Rothmaier, D.~Miralles, D.~Akos, and T.~Walter, ``{Developing a Practical GNSS Spoofing Detection Thresholds for Receiver Power Monitoring},'' in \emph{Proceedings of the 34th International Technical Meeting of the Satellite Division of The Institute of Navigation (ION GNSS+ 2021)}, St. Louis, Missouri, Sep. 2021, pp. 803--815.

\bibitem{Wesson2018}
K.~Wesson, J.~Gross, T.~Humphreys, and B.~Evans, ``Gnss signal authentication via power and distortion monitoring,'' \emph{IEEE Transactions on Aerospace and Electronic Systems}, vol.~54, no.~2, pp. 739--754, 4 2018.

\bibitem{Curran2017}
J.~T. Curran, M.~Bavaro, P.~Closas, and M.~Navarro, ``{A Look at the Threat of Systematic Jamming of GNSS},'' \emph{InsideGNSS}, 2017.

\bibitem{ODriscoll2023}
C.~O'Driscoll, J.~Winkel, and I.~F. Hernandez, ``Assisted {NMA} proof of concept on android smartphones,'' in \emph{2023 {IEEE}/{ION} Position, Location and Navigation Symposium ({PLANS})}.\hskip 1em plus 0.5em minus 0.4em\relax {IEEE}, apr 2023.

\bibitem{McNeill2017}
J.~McNeill, S.~Razavi, K.~Vedula, and D.~R. Brown, ``Experimental characterization and modeling of low-cost oscillators for improved carrier phase synchronization,'' in \emph{2017 {IEEE} International Instrumentation and Measurement Technology Conference (I2MTC)}.\hskip 1em plus 0.5em minus 0.4em\relax {IEEE}, may 2017.

\bibitem{Steingass2004}
A.~Steingass and A.~Lehner, ``Measuring the navigation multipath channel - a statistical analysis,'' in \emph{Proceedings of the 17th International Technical Meeting of the Satellite Division of The Institute of Navigation}, Long Beach, Sep. 2004.

\bibitem{Lehner2007}
A.~Lehner, \emph{Multipath Channel Modelling for Satellite Navigation Systems}, ser. Berichte aus der Luft- und Raumfahrttechnik.\hskip 1em plus 0.5em minus 0.4em\relax Aachen: Shaker Verlag, 2007.

\bibitem{Sanz2017}
\BIBentryALTinterwordspacing
J.~Sanz, J.~Miguel~Juan, A.~Rovira-Garcia, and G.~González-Casado, ``{GPS} differential code biases determination: methodology and analysis,'' \emph{GPS Solutions}, vol.~21, no.~4, pp. 1549--1561, 2017. [Online]. Available: \url{https://doi.org/10.1007/s10291-017-0634-5}
\BIBentrySTDinterwordspacing

\bibitem{Larson2018}
J.~Larson, ``{Gaussian-Pareto} overbounding: A method for managing risk in safety-critical navigation systems,'' phdthesis, University of Minnesota, 2018.

\end{thebibliography}

\end{document}